\documentclass[9pt]{article} 

\usepackage{times} 
\usepackage{graphicx} 
\usepackage[english]{babel}

\usepackage{amssymb}

\newcommand\mean[1]{{\langle#1\rangle}}
\newcommand\Ocal{{\cal O}}
\newcommand\Hcal{{\cal H}}

\title{The Minority Game: an introductory guide}
\author{Esteban Moro\footnote{Grupo Interdisciplinar de Sistemas Complicados (GISC) and
Departamento de Matem\'aticas, Universidad Carlos III de Madrid,
Avenida de la Universidad 30, E-28911 Legan\'es, Spain. email:
{\tt emoro@math.uc3m.es}}}

\begin{document}                      

\date{}

\maketitle

\begin{abstract}
The Minority Game is a simple model for the collective behavior of
agents in an idealized situation where they have to compete
through adaptation for a finite resource. This review summarizes
the statistical mechanics community efforts to clear up and
understand the behavior of this model. Our emphasis is on trying
to derive the underlying effective equations which govern the
dynamics of the original Minority Game, and on making an
interpretation of the results from the point of view of the
statistical mechanics of disordered systems.
\end{abstract}

\section{Introduction}
\begin{flushright}
\begin{minipage}{7cm}
\scriptsize
There are $10^{11}$ stars in the galaxy. That used to
be a huge number. But it's only a hundred billion. It's less than
the national deficit! We used to call them astronomical numbers.
Now we should call them economical numbers.\\
{\em Richard Feynman}
\end{minipage}
\end{flushright}

\vspace{-12cm}{\small \noindent To appear in {\em Advances in
Condensed
Matter and Statistical Physics}\hfill ISBN 1-59033-899-5 \\
Edited by Elka Korutcheva and Rodolfo Cuerno\hfill \copyright 2004
Nova Science Publishers, Inc.}

\vspace{12cm}

During the last years, the statistical mechanics community has
turned its attention to problems in social, biological and
economic sciences. This is due to the fact that in those sciences,
recent research has focused on the emergence of aggregates like
economic institutions, migrations, patterns of behavior, etc., as
a result of the outcome of dynamical interaction of many
individual agent decisions
\cite{anderson1,kosfeld,hpyoung,schelling}. The differences of
this approach with traditional studies are twofold: firstly, the
fact that collective recognizable patterns are due to {\em the
interaction of many individuals.} On the other hand, in the
traditional neoclassical picture, agents are assumed to be
hyperrational and have infinite information about other agents'
intentions. Under these circumstances, individuals jump directly
into the steady state. However individuals are far from being
hyperrational and their behavior often changes over time. Thus,
new models consider explictly {\em the dynamical approach towards
the steady state} through evolution, adaptation and/or learning of
individuals. The dynamical nature of the problem poses new
questions like whether individuals are able to reach a steady
state and under which circumstances this steady state is stable.

The problem is thus very appealing to statistical mechanics
researchers, since it is the study of many interacting degrees of
freedom for which powerful tools and intuitions had been
developed. However, the dynamics of the system might not be
relaxing and the typical energy landscape in which steady states
are identified with minima of a Lyapunov function so that the
dynamical process executes a march which terminates at the bottom
of one of the valleys could not be applicable. The situation is
reminiscent of other areas at the borderline of statistical
mechanics like Neural Networks \cite{amit} where sometimes is not
possible to build up a Lyapunov function which the dynamics tend
to minimize. In fact, as we will see below there are strong
analogies of the model considered here with that area of research.

The model I consider here is called the Minority Game (MG)
\cite{challet1997} which is the mathematical formulation of ``El
Farol Bar'' problem considered by Brian Arthur \cite{arthur94}.
The idea behind this problem is the study of how many individuals
may reach a collective solution to a problem under adaptation of
each one's expectations about the future. As the models mentioned
before, the MG is a dynamical system of many interacting degrees
of freedom. However, the MG includes two new features which make
it different: the first one is the minority rule, which makes a
complete steady state in the community impossible. Thus, dust is
never settled since individuals keep changing and adapting in
quest of a non-existing equilibrium. The second ingredient is that
the collectivity of individuals is heterogeneous, since
individuals have different ways to tackle available information
about the game and convert it into expectations about future.
Effectively, the minority rule and heterogeneity translate into
mean-field interaction, frustration and quenched disorder in the
model, ideas which are somehow familiar to disordered systems in
condensed matter \cite{sherrington99,young}. Actually, the MG is
related to those systems and some of the techniques of disordered
systems can be applied to understand its behavior.

At this point, I  hope the reader is convinced that the MG is an
interesting problem from the statistical mechanics point of view:
is a complex dynamical disordered system which can be understood
with techniques from statistical physics. In fact, most of the
research about the MG problem has been done inside the physics
community. However, the El Farol bar problem originates in the
economic literature, although it represents a fierce assault on
the conventions of standard economics. In this sense either the El
Farol bar or the MG are interesting from the economic and game
theory point of view as well. My intention here is to present the
statistical mechanics approach to this problem and to explain how
different concepts of economists and game theory translate into
physics terminology. This biased review is the author's effort  to
explain to physicists what is known about the MG. I make no
attempt to be encyclopedic or chronologically historical.

\section{The El Farol Bar problem}
\begin{flushright}
\begin{minipage}{9cm}
\scriptsize EL FAROL, 808 Canyon Rd., Santa Fe: El Farol
restaurant is where locals like to hang out. The roomy, somewhat
ramshackle bar area has an old Spanish-western atmosphere, but you
can order some fine Spanish brandies and sherries. On most nights
you can listen to live flamenco, country, folk, or blues. Dancers
pack the floor on weekend nights in summer.\\
{\em Fodor's miniguide}

\end{minipage}
\end{flushright}

The El Farol Bar problem was posed as an example of inductive
reasoning in scenarios of bounded rationality \cite{arthur94}. The
type of rationality which is usually assumed in economics
--perfect, logical, deductive rationality -- is extremely useful
in generating solutions to theoretical problems \cite{gametheory},
but if fails to account for situations in which our rationality is
bounded (because agents can not cope with the complexity of the
situation) or when ignorance about other agents ability and
willingness to apply perfect rationally lead to subjective beliefs
about the situation.  Even in those situations, agents are not
completely irrational: they adjust their behavior based on the
what they think other agents are going to do, and these
expectations are generated endogenously by information about what
other agents have done in the past. On the basis of these
expectations, the agent takes an action, which in turn becomes a
precedent that influences the behavior of future agents. This
creates a feedback loop of the following type:
\begin{center}
\begin{tabular}{rcl}
Precedents & $\longrightarrow$ & Expectations \\
$\nwarrow$ &  & $\swarrow$ \\
& Actions &
\end{tabular}
\end{center}
Inductive reasoning assumes that by feeding back the information
about the game outcome, agents could eventually reach perfect
knowledge about the game and arrive to an steady state. On the
contrary, deductive reasoning (which is commonly applied in
economics, for example), assume that the precedents contain full
information about the game and then there is not any dynamical
approach to the steady state, which is attained in a single step:
\begin{center}
\begin{tabular}{rcl}
Precedents (full information) & $\longrightarrow$ & Actions (steady state)\\
\end{tabular}
\end{center}

The El Farol bar problem is posed in the following way: $N$ people
decide independently each week whether to go to a bar that offers
entertainment on a certain night. Space is limited, and the
evening is enjoyable if things are not too crowed --specifically,
if fewer than $aN$ with $a < 1$ of the possible $N$ are present.
In other case, they stay at home. Thus agents have two actions:
{\em go} if they expect the attendance to be less than $aN$ people
or {\em stay at home} if they expect it will be overcrowded. There
is no collusion or prior communication among the agents; the only
information available is the numbers who came in past weeks. Note
that there is no deductively rational solution to this problem,
since given only the numbers attending in the recent past, a large
number of expectational models might be reasonable. Thus, not
knowing which model other agents might choose, a reference agent
cannot choose his in a well-defined way. On the other hand,
expectations are forced to differ: if all believe most will go,
nobody will go, invalidating that belief. In order to advance the
attendance next week each agent is given a fixed number of {\em
  predictors} which map the past weeks' attendance figures into
next week. The idea behind those predictors is that when agents face
complex scenarios like this they tend to look for patterns that
occurred in the past and predict next move from that experience.
Finally, each agent monitors his predictors by keeping an internal
score of them which is updated each week by giving points or not to
all of them whether they predicted the attendance or not. At each
weak, each agent uses his predictor with bigger score. Computer
simulations of this model \cite{arthur94} shows that the attendance
fluctuates around $aN$ in a ($aN$,$(1-a)N$) structure of people
attending/not attending. Thus, predictors self-organize so that this
structure emerges in the complex dynamics of the system.

Despite the fact that the El Farol bar problem deals apparently
with a non-market context it can be considered as a kind of very
simple ``proto-market'' or ``market toy model''
\cite{manucaphysa2000,marsili_review,zhang}: at each time step
agents can buy (go to the bar) or sell an asset. After each time
step, the price of the asset is determined by a simple
supply-demand rule: if there are more buyers than sellers, the
price is high, and if there are more sellers than buyers, the
price is low. If the price is high, sellers do well, while if the
price is low, buyers win the round. Thus, the minority group
always win, irrespectively of whether they were buyers or sellers.
However, some of the behavioral assumptions on which the El Farol
bar problem (and the MG model) is based may be questionable when
applied to financial markets \cite{cmzphysa2000}. Moreover, it
still lacks some of the most basic features in a real market,
e.g.\ the price which is determined by the aggregate supply-demand
rule is not considered or the fact that agents do have different
payoffs for their actions is not taken into account \cite{zhang}.
The relation to the markets is then on the conceptual level, as
markets are interpreted as an adaptative competitive system in
which minority membership plays an important role. Many other
complex systems are driven by this minority rule, like vehicular
traffic on roads, in which each agent would prefer to be on an
uncongested road \cite{nagel1}, packet traffic in networks, in
which each packet will travel faster through lesser used routers
\cite{huberman} and ecologies of animals looking for food, in
which individuals do best when they can find areas with few
competitors \cite{macara,deangelis}.

From the statistical mechanics point of view, we can say that in
the El Farol bar problem, a system with $N$ degrees of freedom has
finally come to a stationary state in which the attendance
fluctuates around $aN$. Some of the properties of the El Farol bar
problem are very appealing to statistical mechanics, like:
\begin{itemize}
\item Many ($N \gg 1$) interacting agents. \item Interaction is
through the aggregate bar attendance, i.e. of the  mean field
type. \item The system of $N$ agents is frustrated, since there is
not a unique
  winning strategy in the problem.
\item Quenched disorder, since agents use different predictors to
generate expectations about the future of the game.
\end{itemize}
However, other features of the model are not usual in statistical
mechanics, like the non-Markovian time dynamics induced by the
fact that the predictors map past attendance into next week
prediction which is released back to the agents. Moreover, the
dynamics based on inductive reasoning may not be relaxing and, in
principle, the steady state which the game attains could be out of
equilibrium: there is no reason {\em a priori} to expect detailed
balance or ergodicity in the system dynamics. At this stage it is
not clear which of the main ingredients outlined before is
relevant for the observed behavior. Thus, it is necessary first to
give a precise mathematical definition of the model and try to
clear up what are the ingredients truly responsible for the model
main features.

\section{The Minority Game}\label{sec_mg}
\begin{flushright}
\begin{minipage}{9cm}
\scriptsize
Truth always rests with the minority, and the minority is
always stronger than the majority, because the minority is generally
formed by those who really have an opinion, while the strength of a
majority is illusory, formed by the gangs who have no opinion--and who,
therefore, in the next instant (when it is evident that the minority is
the stronger) assume its opinion . . . while Truth again reverts to a
new minority.\\
{\em Soren Kierkegaard}
\end{minipage}
\end{flushright}

In order to rationalize the discussion, Challet and Zhang
\cite{challet1997} gave a precise mathematical definition for the
El Farol bar problem which they called the Minority Game (MG). In
the MG the predictors are replaced by well defined strategies from
a given state space and the available information used by the
agents is accurately defined.

\begin{figure}
\begin{center}
\includegraphics[width=0.3\textwidth]{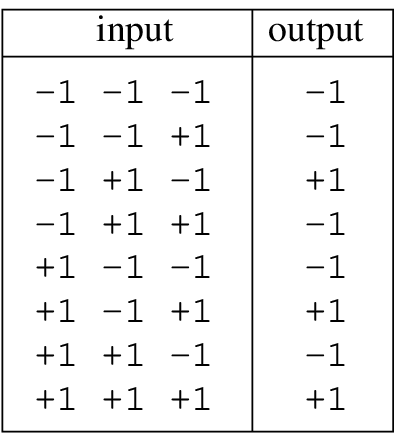}\ \ \ \ \
\includegraphics[height=0.4\textwidth]{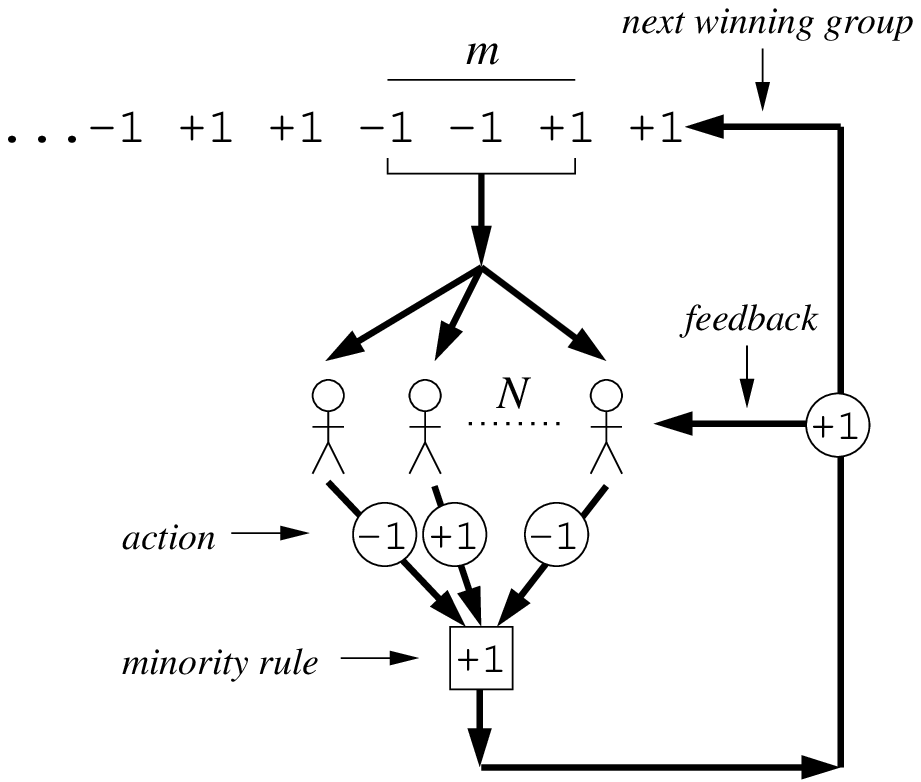}
\caption{{\em Left}: An example of an strategy for $m=3$. {\em
Right}: Cartoon of the MG model for a given time step: in this
case the strategy maps the last three winning groups ($m=3$) into
the agent decision. Solid thick lines mimic how the information
flows in the system: the $N$ agents take the last $m$ numbers
($-1,-1,+1$ in this case) from the sequence of winning groups and
perform an action accordingly. The $N$ actions are transformed
into the next winning group ($+1$ in this case) through the
minority rule. This information is shared with the agents for
their own feedback [Eq. (\ref{eq_points})] and becomes the next
number in the sequence of winning groups.} \label{fig_tabla}
\end{center}
\end{figure}

In their definition of the MG model, the game consists of $N$
(odd) agents playing as follows (see figure \ref{fig_tabla}): at
each time step of the game, each of the $N$ agents takes an action
$a_i(t)$ with $i=1,\ldots,N$: he decides either to go to the bar
($a_i(t) = 1$) or to stay at home ($a_i(t)=-1$). The agents who
take the minority action win, whereas the majority looses. After a
round, the total action is calculated
\begin{equation}\label{attendance}
A(t) = \sum_{i=1}^N a_i(t).
\end{equation}
The minority rule sets the comfort level at $A(t) = 0$, so that
agent is given a payoff  $-a_i(t) g[A(t)]$ at each time step with
$g$ an odd function of $A(t)$. Challet and Zhang's initial choice
was $g(x) = \mathrm{sign} (x)$, but other analytical functions are
more suitable for mathematical treatment like $g(x) = x/N$. Most
of the MG properties are qualitatively independent of the precise
analytical form of $g(x)$ \cite{li2000}. The information about the
winning group is released to the agents which is given by $W(t+1)
= \mathrm{sign} A(t)$.

The way agents choose their action $a_i(t)$ is by inductive
reasoning: we assume that the players have limited analyzing power
and they can only retain the last $m$ winning groups. Moreover
they base their decision $a_i(t)$ on these last $m$ bits only. To
this end they have a set of $s \geq 2$ strategies. A strategy is
just a mapping from the sequence of the last $m$ winning groups to
the action of the agent, i.e. a table which tells the agent what
to do as a function of the input signal (the last $m$ winning
groups). An instance of one of those strategies is given in figure
\ref{fig_tabla}.

Since there are $2^m$ possible inputs for each strategy, the total
number of possible strategies for a given $m$ is $2^{2^m}$. At the
beginning of the game each agent is given a set of $s$ strategies
randomly drawn from the total $2^{2^m}$ possible strategies. This
assignment is different for each agent and thus, agents may or may
not share the same set of strategies. Note that each strategy is
determined by a $2^m$ dimensional vector $\vec{r}^{\:\alpha}_i$
whose components are the output of strategy $\alpha$ of agent $i$.
For example, in figure \ref{fig_tabla} the strategy is
$\vec{r}^{\:\alpha}_i = (-1,-1,+1,-1,-1,+1,-1,+1)$. If the last
winning groups were $-1, +1, +1$, then the prediction for this
strategy is given by the fourth component of $\vec{r}^{\:\alpha}$.
Thus, at each time step, the prediction of a strategy is given by
its $\mu(t) \in \{1,\ldots,2^m\}$ component, where $\mu(t)$ is a
number whose binary representation correspond to the last winning
groups\footnote{In the binary representation of $\mu(t)$ we make
the correspondence $-1 \leftrightarrow 0$ and $+1 \leftrightarrow
1$. Thus, if the last winning groups were $-1,+1,+1$ the binary
representation of $\mu(t)$ is $011$ and $\mu(t) = 4$.}. If we
denote $\vec{I}(t)$ the vector whose components are zero excepting
the $\mu(t)$ component which is one, then the prediction of
strategy $\vec{r}_i^{\:\alpha}$ is given by $\vec{r}_i^{\:\alpha}
\cdot \vec{I}(t)$. For example, if the last three winning groups
were $-1,+1,+1$ then $\mu(t) = 4$ and $\vec{I}(t) =
(0,0,0,1,0,0,0)$. Thus, strategies are all possible
$2^m$-dimensional vectors with $\pm 1$ components.

Adaptation comes in the way agents choose at each time step one of
their $s$ strategies: they take the strategy within their own set
of strategies whose performance over time to predict the next
winning group is biggest. In order to do that each agent $i$
assigns virtual points $p_i^{\alpha}(t)$ to his strategy $\alpha$
after each time step $t$ when they predicted correctly the winning
group:
\begin{equation}\label{eq_points}
p_i^\alpha(t+1) = p_i^\alpha(t) - \vec{r}_i^{\:\alpha}\cdot\vec{I}(t)\ g[A(t)]
\end{equation}
where $\alpha = 1,\ldots,s$ and $i = 1,\ldots, N$. However these
points are only virtual points as they record only agents'
strategies performance and serve only to rank strategies within
each agent set. After time $t$ agent $i$ takes the first strategy
in his personal ranking which tells him what to do in the future.
If we denote agent $i$ best strategy at time $t$ in his ranking as
$\beta_i(t) \in \{1,\ldots,s\}$,\footnote{When two strategies have
the highest number of points, the best strategy is chosen by coin
tossing.} then his action at time $t$ is given by:
\begin{equation}\label{eq_action}
a_i(t) = \vec{r}_i^{\:\beta_i(t)}\cdot \vec{I}(t).
\end{equation}
The fact that this personal ranking can change over time makes the
agents adaptative: the ranking of each agent's strategies can
change over time and then $\beta_i(t)$ could be different at
different times.

Finally, the information $\vec{I}(t)$ is updated by adding the
last winning group: the only nonzero component of $\vec{I}(t+1)$
is the $\mu(t+1)$ one which is given by \cite{challetmarsilipre99}
\begin{eqnarray}
\mu(t+1) = [2\mu(t) + (W(t+1)-1)/2] \mathrm{mod} P,
\label{eq_information}
\end{eqnarray}
were $W(t+1) = \mathrm{sign} A(t)$ is the winning group.

Equations (\ref{eq_points}) and (\ref{eq_action}) together with
the implicit ranking of each agent's strategies define the process
of adaptation. Note that the minority rule is encoded in the
function $g(x)$ of the attendance and appears in the way public
information $\vec{I}(t)$ is built and also when virtual points are
given to the strategies. Finally, interaction between agents in
eq.\ (\ref{eq_points}) is through the attendance $A(t)$ which is
of the mean-field type. The heterogeneity among agents shows up at
the set of $\vec{r}_i^{\:\alpha}$ which could be different for
different agents.

An interesting point is to observe that the variables which define
the state of the game at each time step are $p_i^{\alpha}(t)$
together with the public information $\vec{I}(t)$. But, on the
other hand, eq.\ (\ref{eq_information}) and (\ref{eq_points})
introduce a non trivial feedback in the dynamics since the state
of the system at time $t$ depends on the last $m$ winning groups.
This is one of the characteristic features of the MG, since the
dynamics given by eqs.\ (\ref{eq_points})-(\ref{eq_information})
is non-local in time.

\section{Coordination due to adaptation}\label{sec_adaptation}
\begin{flushright}
\begin{minipage}{9cm}
\scriptsize It is not worth an intelligent man's time to be in the
majority. By definition, there are already enough
people to do that.\\
{\em Godfrey Harold Hardy}
\end{minipage}
\end{flushright}

\begin{figure}
\begin{center}
\includegraphics[height=0.42\textwidth]{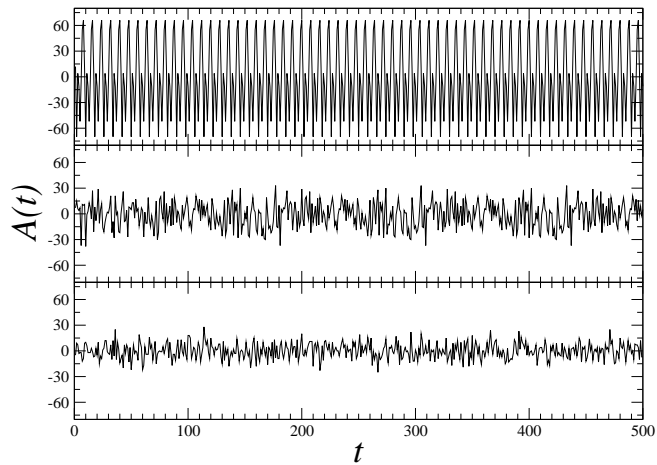}
\caption{Time evolution of the attendance for the original MG with
$g(x) = x$ and $N=301$ and $s=2$. Panels correspond to $m=2, 7$
and $m=15$ from top to bottom. Periodic patterns can be observed
for $m=2$ and $m=7$.} \label{fig_simulations}
\end{center}
\end{figure}

Initial studies of the MG model relied in simulations
\cite{macara1,challet1997,challet1998,manucaphysa2000,johnson1}.
Typical simulations of the MG in its original formulation $g(x) =
\mathrm{sign}(x)$ are given in figure \ref{fig_simulations}. As we
mentioned in the introduction, the aggregate $A(t)$ never settles
down and it fluctuates around the comfort level, $A(t) = 0$, as
observed by Arthur in his paper \cite{arthur94}. Thus we have
$\overline{\mean{A(t)}} = 0$ where $\mean{\cdots}$ is a time
average for long times and $\overline{\cdots}$ is an average over
possible realizations of $\vec{r}_i^{\:\alpha}$. Despite its
trivial mean value the possible values of $A(t)$ display a
nontrivial shape and the fluctuations are important
\cite{macara1}. Moreover for small values of $m$ the attendance
display time periodic patterns which are lost for large values of
$m$.

\subsection{Volatility}
While the behavior of $\overline{\mean{A(t)}}$ is somehow trivial,
fluctuations of $A(t)$ around its mean value given by the variance
$\sigma^2 = \overline{\mean{[A(t)-\mean{A(t)}]^2}}$ have a more
interesting behavior (see figure \ref{figoriginal}). First note
that $\sigma$ is related to the typical size of the losing group,
so the smaller $\sigma$, the more winners are in the game. The
variance $\sigma^2$ is usually known as the volatility or (the
inverse of) global efficiency. The behavior of $\sigma^2$ as a
function of the parameters of the model $m$, $s$ and $N$ shows a
quite remarkable behavior \cite{manucaphysa2000,savitprl1999}:
\begin{figure}
\begin{center}
\includegraphics[height=0.3333\textheight]{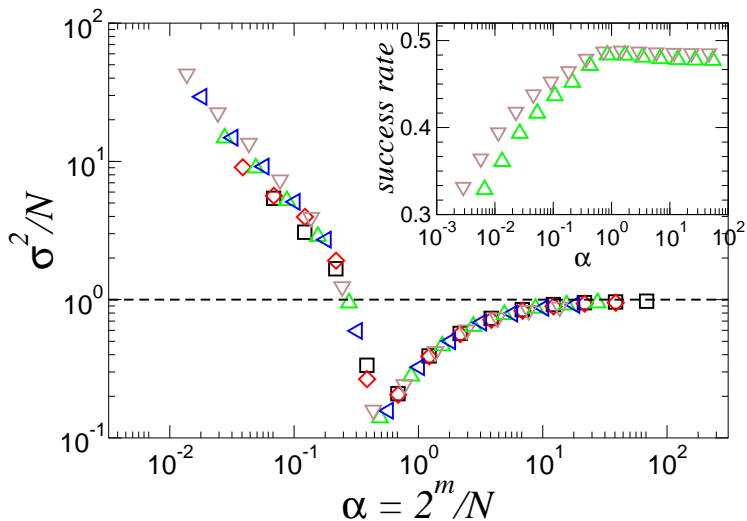}
\caption{Volatility as a function of the control parameter $\alpha
= 2^m/N$ for $s=2$ and different number of agents
$N=101,201,301,501,701$ ($\square$, $\lozenge$, $\triangle$,
$\lhd$, $\triangledown$, respectively). Inset: Agent's mean
success rate as function of $\alpha$.  } \label{figoriginal}
\end{center}
\end{figure}
\begin{itemize}
\item It was found by extensive simulations that $\sigma^2/N$ is
only a function of $\alpha = 2^m/N$ for each value of $s$ (see
figure \ref{figoriginal}). This finding not only identifies the
control parameter in this model, $\alpha$, but also paves the way
for the application of tools of statistical mechanics in the
thermodynamic limit $N\to \infty$. Since qualitative results are
independent of $s \geq 2$ we take the simplest case $s=2$ for the
rest of the chapter. \item For large values of $\alpha$,
$\sigma^2/N$ approaches the value for the random choice game
$\sigma^2/N =1$, i.e., the game in which each agent randomly
chooses $a_i(t) = -1$ or $a_i(t) = 1$ independently and with equal
probability at each time step. \item At low values of $\alpha$,
the average value of $\sigma^2$ is very large, actually, it scales
like $\sigma^2/N \sim \alpha^{-1}$ which means that $\sigma \sim
N$ and thus the size of the losing group is much larger than
$N/2$. \item At intermediate values of $\alpha$, the volatility
$\sigma$ is less than the random case, and it attains a minimum
value at $\alpha \simeq 1/2$. In this region, the size of the
losing group is close to $N/2$ (which is the minimum possible size
for the losing group).
\end{itemize}

The fact that $\sigma$ gets below the random case for a given
interval of values of $\alpha$ suggests the possibility that
agents coordinate in order to reach a state in which less
resources are globally wasted \cite{savitprl1999}. In the market
formalism, this means that agents can exploit information
available and predict future market movements, so that global
efficiency $\sigma^2$ is minimized. In this sense {\em adaptation}
is a good mechanism for the agents to achieve a better solution to
the MG problem for $\alpha =\alpha_c \simeq 1/2$. The solution
comes through coordination among agents, which is contrary to
their selfish intrinsic nature. Global coordination as the better
solution to a problem of competition for resources is common in
different games and situations \cite{gametheory}.

Note however that in the MG coordination is not complete and the
best possible solution is not achieved: that in which agents
alternate in groups of $(N-1)/2$ and $(N+1)/2$ which yields
$\sigma^2/N = 1/N$. In this situation the mean success rate, i.e.\
the frequency an agent is successful in joining the minority is
$1/2$. As shown in figure \ref{figoriginal} the success rate in
the MG is close to $1/2$ around the minimum of $\sigma^2/N$, while
it is substantially smaller in the region where the volatility is
high.

\subsection{Information}

\begin{figure}
\begin{center}
\includegraphics[height=0.45\textwidth,clip=]{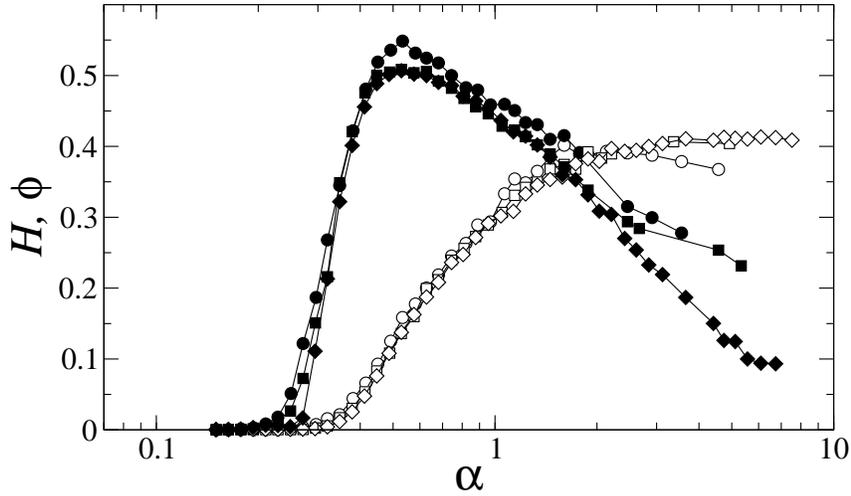}
\caption{Information $H$ (open symbols) and fraction of frozen
agents $\phi$ (full symbols) as a function of the control
parameter $\alpha = 2^m/N$ for $s=2$ and $m = 5,6,7$ (circles,
squares and diamonds respectively).} \label{fig-information}
\end{center}
\end{figure}

The fact that agents seem to exploit the available information led
some authors to study the information contained in the time series
of $A(t)$. Specifically, it was found that $W(t+1) = \mathrm{sign}
A(t)$ is independent of sequence of the last $m$ attendances in
the high volatility region, while there is a strong dependence for
$\alpha > \alpha_c$. From the market point of view, the region
where $\alpha < \alpha_c$ is called efficient (although is
socially inefficient), since the history of minority groups
contains no predictive information for agents with memory $m$. On
the contrary, in the region where $\alpha > \alpha_c$ there is
significant information available to the strategies of the agents
playing the game with memory $m$ and the market is not efficient
in this sense, since there are arbitrage
possibilities\footnote{The existence of arbitrage opportunities in
a market implies that there exist price discrepancies of the same
stock which can be exploit by agents to make profit by selling and
buying them simultaneously. The market is efficient whenever these
arbitrage opportunities are exploited instantaneously and
therefore eliminated from the market.}. But in either case $A(t)$
is not a random sequence (see figure \ref{fig_simulations})
\cite{macara1,manucaphysa2000}. To quantify this behavior, it was
proposed in \cite{manucaphysa2000} to measure the conditional
probability of $W(t+1)$ knowing $\mu(t)$ through the mutual
entropic information of $W(t)$ and $W(t+1)$. Another more direct
measure of this quantity was suggested in
\cite{challetmarsilipre99} as
\begin{equation}\label{def_information}
H = \frac{1}{2^m} \sum_{\nu = 1}^{2^m} \mean{W(t+1) | \mu(t) = \nu}^2
\end{equation}
where the average of $W(t+1)$ in (\ref{def_information}) is
conditioned to the requirement that the last $m$ winning groups
are given by $\mu(t)$. If there is no significant dependence
between $W(t+1)$ and $\mu(t)$ then as we have already seen
$\mean{W(t+1)} = \mean{\mathrm{sign} A(t)}=0$ and $H=0$. Loosely
speaking $H$ measures the information in the time series of $A(t)$
available to the agents. In the market context, $H\neq 0$
indicates the presence of information or arbitrages in the signal
$A(t)$. As seen in figure \ref{fig-information}, $H = 0$ for
$\alpha < \alpha_c \simeq 0.3$ while $H \neq 0$ for $\alpha
> \alpha_c$. When $\alpha \to \infty$ we recover $H=0$ since in
that case the system of $N$ agents behave almost randomly and we
expect no correlations in the sequence of $W(t)$.

Despite the fact that $H$ measures the information content in the
sequence of $A(t)$ it does it only through the correlation of the
next winning group $W(t+1)$ with the last $m$ winning groups.
Moreover, $H$ also measures the asymmetry of the {\em response} of
the agents to the information available to them: when $H \neq 0$
the system of $N$ agents reacts differently to the available
information, while when $H=0$ all given information is treated
equally by the agents. Thus, $H$ has a twofold role in the game as
a measure of the information available in the sequence of $A(t)$
and the response of the system to a given piece of information
$\mu(t)$.

\subsection{Phase transition}

The fact that $H=0$ for $\alpha < \alpha_c$ and $H\neq 0$ for
$\alpha > \alpha_c \simeq 0.3$ suggests the possibility that there
is a phase transition at $\alpha = \alpha_c$ which separates those
two efficient and inefficient phases \cite{challetmarsilipre99}.
Thus $H$ is the order parameter of the system which measures the
symmetry of $W(t+1)$ in the system, a symmetry which is broken at
$\alpha = \alpha_c$. Since the phase transition occurs at $\alpha
= \alpha_c$, it means that for a fixed value of $m$, the agents
who enter the market exploit the $A(t)$ predictability and hence
reduce it. From the point of view of the community, $\alpha_c$
separates two regions in which adaptation is successful in
achieving a good global solution to the MG problem. The following
table summarizes the different names for the phases given in
different contexts:

\bigskip

\begin{center}
\begin{tabular}{|c|c|c|}
\hline
Context & $\alpha < \alpha_c$ & $\alpha > \alpha_c$ \\
\hline
Volatility/  & Inefficient & Efficient \\
Global Waste &  (Worse than random) & (Better than random) \\
\hline
Information/ & Efficient & Inefficient \\
Arbitrage in $A(t)$ & (no information, $H=0$) & (arbitrage, $H\neq 0$).\\
\hline
\end{tabular}
\end{center}

\bigskip

Note that the phase transition is both explained in terms of the
information which is available to the agents and the minimization
of the total waste, while only $H$ is an order parameter. Other
quantities have been identified as order parameters to explain the
phase transition found numerically at $\alpha = \alpha_c$. For
example in \cite{challetmarsilipre99} it was considered the
fraction of agents $\phi$ who always use the same strategy, i.e.
those agents who have $\beta_i(t) = \beta_i$, independently of
time. As we see in figure \ref{fig-information} $\phi$ is zero in
the $\alpha < \alpha_c$ phase which means that all the agents
change their strategy at least once. This is also the case for
large values of $\alpha$, since agents behave more or less
randomly. However close to the transition point the fraction
$\phi$ is large and most of the agents stick to one of their
strategies forever. In terms of $\phi$ the phase transition can be
understood as an ``unbinding'' transition, since for $\alpha <
\alpha_c$ strategy points $p_i^{\alpha}(t)$ remain very close to
each other and then agents can switch easily from one strategy to
the other. However, for $\alpha > \alpha_c$ one of each agent's
strategies wins in the long run and its virtual score
$p_i^{\alpha}(t)$ is much larger than the others.

Finally, while there is no significant finite size corrections to
the order parameter $H$, the fraction of frozen agents $\phi$
seems to drop discontinuously to zero at $\alpha_c$ when the
number of agents is changed. Despite both order parameters
indicate that the system undergoes a phase transition at $\alpha =
\alpha_c$ it is not yet clear which is the nature of the phase
transition (first, second order?) and whether it is an equilibrium
or non-equilibrium phase transition.

\section{Spanning the strategy space}\label{section_spanning}
\begin{flushright}
\begin{minipage}{6cm}
\scriptsize
Nobody goes there anymore. It's too crowded.\\
{\em Yogi Berra  1925-, American Baseball Player}
\end{minipage}
\end{flushright}

The universality of the behavior found in the last section for
different number of agents, strategies or memory, suggests that it
should depend on a generic feature of the game. Due to the
minority rule, agents tend to differentiate: if they use the same
strategy then all of them will lose. Thus, somehow the minority
rule forces the agents to choose that strategy among theirs that
makes them singular in the community. This translates into a
repulsion of agents chosen strategy inside the strategy space. But
the set of available strategies only contains $2^{2^m}$
strategies. Thus one should expect that agents are very different
between them if the number of strategies is bigger than $N$, while
they should behave more like a crowd if their strategy
heterogeneity is small. In other words we should observe a change
in the system behavior when $N \sim 2^{2^m}$. However as we saw in
the last section this change of behavior is observed at $N \sim
{2^m}$, i.e.\ $\alpha = \Ocal(1)$.

\begin{figure}
\begin{center}
\includegraphics[width=0.66\textwidth,clip=]{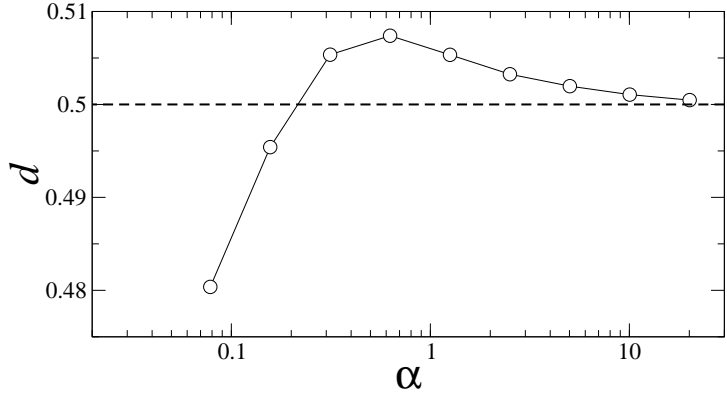}
\caption{Distance of agents' best strategies as a function of
$\alpha$. Parameters used are $N=51$, $s=2$} \label{distancia}
\end{center}
\end{figure}

As note by various authors, while $2^{2^m}$ is the number of
possible strategies, $2^m$ is the dimension of the space in which
strategies live and thus there are only $2^m$ completely different
strategies \cite{challet1998,johnsoncrowd,johnson1}\footnote{Two
strategies are said to be completely different if none of their
components match. A quantitative measure of this is the Hamming
distance, see Eq.\ (\ref{def_distance}).}. If agents tend to
differentiate they can only do it by choosing those completely
different strategies and this is only possible if $N < 2^m$. For
$N > 2^m$ groups of agents with the same set of similar strategies
emerge. Those agents end up using their same best strategy and
become a {\em crowd}, since they react to the available
information in the same way. The different behavior of
$\sigma^2/N$ can be understood in terms of this crowds
\cite{johnsoncrowd,johnson1}:
\begin{itemize}
\item When $N < 2^m$, agents carry a considerable fraction of all
possible strategies. This means that for a large fraction of
agents their best strategy $\beta_i(t)$ is the same at time $t$.
Then the size of this crowd is of the order of $\Ocal(N)$, which
means that the variance is of order $\sigma^2/N \sim N$.
\item On
the other hand, when $N \ll 2^m$ the crowds will be low populated
since the $\beta_i(t)$ are almost different for any agent due to
the high heterogeneity, which makes agents act independently.
Thus, $\sigma^2/N \sim 1$.
\item Finally, for moderate values of
$\alpha$ crowds form which share $\beta_i(t)$ and with size of
order of $N$. But at the same time there will be an {\em
anti-crowd} which is a group of people which have the same best
strategy but which is completely different to the one the crowd is
using. Note that only at moderate values of $\alpha$ this is
possible, since there is a nonzero probability that two groups of
agents cannot share any of their strategies. In this case crowds
and their anticrowds do completely the opposite: considering the
extreme case in which the size of the crowd is $N/2$ we have that
$\sigma^2/N \simeq 0$.
\end{itemize}
More quantitative approximations to the size of the crowds and
anticrowds are possible which reproduce the observed volatility
behavior \cite{johnson1}. Another way to characterize the
formation of crowds and/or anticrowds is to measure the distance
between the best strategies of all the agents \cite{challet1998}
\begin{equation}\label{def_distance}
d = \frac{1}{2^m (N-1)^2}\sum_{i\neq
j}\mean{||\vec{r}_i^{\:\beta_i(t)} -\vec{r}_j^{\:\beta_j(t)}||_1}
\end{equation}
where $||\cdot||_1$ is the Hamming distance (1-norm) in the
$2^m$-dimensional space of the strategies. As seen in figure
\ref{distancia}, agents tend to differentiate between them and
thus maximizing the distances between their best strategies.
However, this is not possible for small values of $\alpha$, since
then agents are force by their limited memory to share their
strategies. The situation is better for large $\alpha$, since
agents' set of strategies do not overlap and then the distance
between them is random $d \simeq 1/2$ \cite{dhulstrodgers99}.
However, for moderate values of $\alpha$ differentiation makes the
system of agents to organize in groups with completely different
best strategies. In the case the crowds and anticrowds are of size
$(N-1)/2$ and $(N+1)/2$ respectively, and knowing that the
distance between two completely different strategies is $2^m$ we
get $d = N/[2(N-1)]$ which is bigger than the random case.

In summary, the observed behavior in the MG is the result of the
minority rule in the strategy space: the minority rule imposes a
repulsion of agents' chosen strategies at each time. However, the
volume of the strategy space is fixed by $m$ and thus repulsion is
only effective when the dimension $2^m$ is much bigger than the
number of strategies present in the game. Note that at
intermediate values of $\alpha$, repulsion of best strategies
leads to coordination, since the optimal way to achieve
differentiation is by joining an existing crowd or anticrowd.

\section{Inventing the future}\label{sec_information}
\begin{flushright}
\begin{minipage}{5.6cm}
\scriptsize
The best way to predict the future is to invent it.\\
{\em Alan Kay}
\end{minipage}
\end{flushright}

One of the main characteristics of the MG model is that agents
adapt and learn from their past history to achieve a better
personal record. As we saw in the previous section, the memory of
the agents is crucial, because it determines the strategy space
dimension and also because the memory is associated with the
ability of agents to spot patterns in the available information so
they can coordinate better. Somehow, the sequence of the winning
groups contains information about the strategies of all the agents
and having more memory gives an agent the possibility to exploit
this information. For example, for a fixed number of agents $N$,
increasing their memory $m$ could lead to better coordination and
higher individual success. In the limit $\alpha \gg 1 $ however,
more memory implies increasing complexity and random behavior.

\begin{figure}
\begin{center}
\includegraphics[width=0.48\textwidth,clip=]{fig-cavagna}\ \ \
\includegraphics[width=0.48\textwidth,clip=]{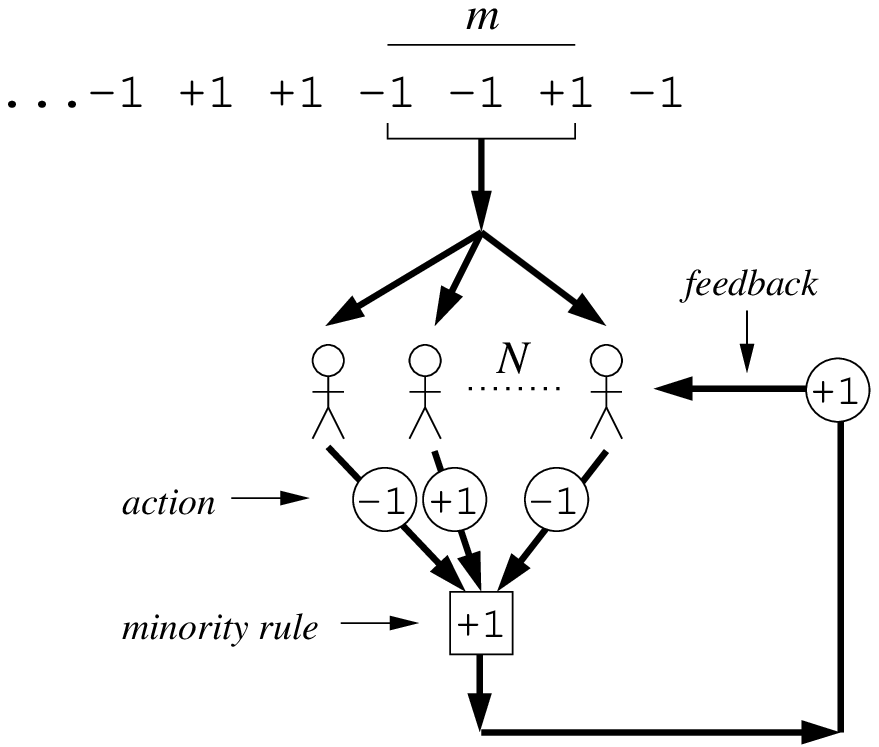}
\caption{{\em Left:} Comparison of the model variance $\sigma^2$,
information $H$ (squares) and $\phi$ (circles) with endogenous
information (open symbols) and exogenous information (full
symbols). {\em Right}: Cartoon of the MG model using exogenous
information. As in figure \ref{fig_tabla} solid thick lines mimic
the flow of information in the system. In this case the winning
group obtained through the minority rule is only used by the
agents to rank their strategies and may differ from the one in the
exogenous sequence of winning groups (as in the figure). }
\label{figcartoon1}
\end{center}
\end{figure}

Since information is fed back into the system, early studies of
the MG concentrated in the possibility that the system of $N$
agents could exploit this information to achieve better
coordination. However, A.\ Cavagna in \cite{cavagna1} showed by
means of extensive simulations that if the information $\mu(t)$
which is given to the agents at each time step is chosen randomly
and independently of time from its possible values $\mu(t) \in
\{1,\ldots,2^m\}$ the behavior of the MG remains the same
regarding time averaged extensive quantities like the volatility,
the information $H$ or $\phi$ (see Fig.\ \ref{figcartoon1}). The
crucial point is that every agent reacts to the same piece of
information whether this information is endogenous or exogenous,
false or true. As a consequence, there is no room in this model
for any kind of forecasting of the future based on the
understanding of the past: coordination in the MG does not come
through exploitation of available information in the sequence of
the winning groups.

Leaving aside the fact that changing the real sequence of $\mu(t)$
for random numbers gives the same quantitative behavior for
different quantities, this finding has significant importance for
the theoretical understanding of the MG model. Specifically,
equation (\ref{eq_information}) drops from the definition of the
model which is simply given by equations (\ref{eq_points}) and
(\ref{eq_action}) together with the ranking of each agent's
strategies at any time step. In this sense, this results are
useful in the theoretical understanding of the MG model, since it
shows that the complicated feedback of the endogenous information
in the system is just an irrelevant complication and thus, can be
dropped in the way to explain the rich structure of this
adaptative model.

This fact does not mean that the memory parameter $m$ is not
relevant in the model: indeed, the model displays a phase
transition at a given value of $m$ for a fixed number of agents.
However, from the results of A.\ Cavagna in \cite{cavagna1} it is
reasonable to conclude that the parameter $m$ only relates to the
dimension $D = 2^m$ of the strategies space. Indeed, as we have
just shown in section \ref{section_spanning}, the description of
the phase transition observed in the MG can be just based on
geometrical considerations about the distribution of strategies in
the space of strategies.

As we can see in figure \ref{figcartoon1} the order parameter $H$
does not change substantially when exogenous information is
considered instead of endogenous one. But, as we said in section
\ref{sec_adaptation}, $H$ is related to correlations in the
sequence of winning groups. Since the sequence of the winning
groups is exogenous and random we expect in $H=0$ for all values
of $\alpha$. The fact that this is not true is due to the twofold
role that $H$ plays in the original MG: $H$ is the response of the
system to a given piece of information and does not depend on the
origin of the information. But (see figure \ref{figcartoon1}):
\begin{itemize}
\item When endogenous information is considered, the response of
the system to the available information does show up in the
sequence of winning groups.
\item When exogenous information is
considered, the response of the system is only considered to get
the winning group and to keep record of the performance of the
strategies and is not included in the sequence of the winning
group.
\end{itemize}
In either case, $H\neq 0$ means that for a given piece of
information, there is a possibility for agents to predict the
response of the system. In this sense, memory is not irrelevant in
the MG (see the discussion about this in
\cite{cavagna1,cavagna2,savitprl2000}).

\begin{figure}
\begin{center}
\includegraphics[width=0.66\textwidth,clip=]{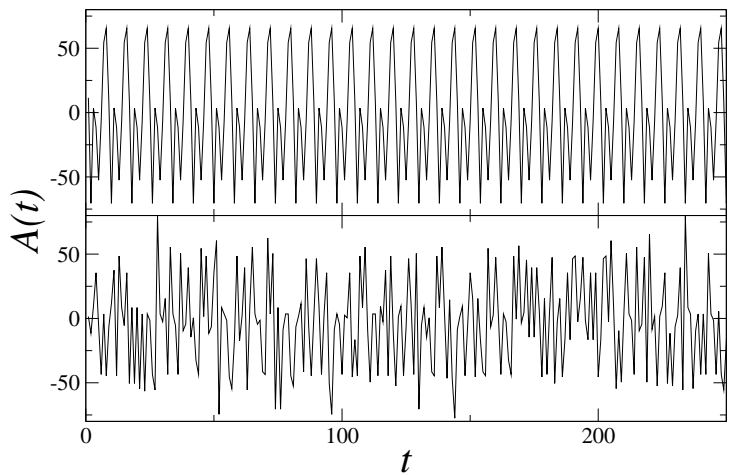}
\caption{Comparison of the bar attendance in the symmetric phase
with endogenous (upper panel) and exogenous information (lower
panel). Parameters used are $N=301$, $m=2$, $s=2$.}
\label{fig_exogenous}
\end{center}
\end{figure}

Despite the fact that the phase transition and macroscopic
properties of the MG are the same with exogenous information, some
properties of the MG are lost. As we show in figure
\ref{fig_simulations} the attendance and the size of the minority
groups are correlated in time and display time periodic patterns
for $\alpha < \alpha_c$. This is lost when the information agents
react to is exogenous as seen in figure \ref{fig_exogenous},
although the typical size of the minority groups is the same and
then $\sigma^2/N$ is independent on the nature of the information
used. Thus, any extension of the MG game which depends on the
periodic nature of the bar attendance should give different
results when exogenous information is considered
\cite{challetmemory,lee,zheng}.

The fact that the memory parameter $m$ is only relevant to
determine the dimension of the strategies space led to the
generalization of equation (\ref{eq_points}) to consider general
$D$-dimensional strategies and vectors $\vec{I}(t)$ for any value
of $D$ and not only for those with $D=2^m$, which is only an
unnecessary complication \cite{cavagna3}.

\section{Maximizing the audience}\label{sec_maximizing}
In section \ref{sec_adaptation} we saw that agents adapt to
achieve a personal better solution to the MG problem which in turn
makes the total waste smaller than the random solution for $\alpha
> \alpha_c$. Thus, although agents are selfish and tend to place
themselves into the minority group irrespectively of other agent's
action, their dynamics tend to maximize the global efficiency,
that is, to maximize the bar attendance. The concept that the
system is trying to minimize a given quantity is very appealing to
statistical mechanics. If this quantity exists, the system can be
studied by considering its minima and perturbations around them.

However, due to the minority rule, the MG never settles down:
specifically, some of the agents keep on changing their strategies
forever and nor $\beta_i(t)$, neither $p_i^{\alpha}(t)$ come to a
rest\footnote{This can be observed in the order parameter $\phi$,
which is never equal to one.}. Thus, there can not be any quantity
based on these degrees of freedom that the MG dynamics tend to
minimize. However, since the actions of the agents $a_i(t)$ depend
on the points of the strategies and $p_i^{\alpha}(t)$ depend on
the past history of actions, one can wonder whether there is any
time pattern in the long run that agents follow. Specifically, it
might be that $a_i(t)$ never come to a rest, but $m_i(t) =
\sum_{\tau=0}^{t} a_i(\tau)$ can converge when $t\to \infty$ to a
given quantity if agent $i$ does follow any behavioral pattern.
For example, agent $i$ can be within the fraction $\phi$ of people
who always chooses one of his strategies and then $m_i(t)$ is the
average of the possible outcomes of that strategy for all possible
realizations of the information. This is the key point in the
first attempt at solution of the MG model found in
\cite{cmzprl2000,mczphysa2000}, which we outline in this section.

First of all, let us rewrite the equations of the MG at this
stage. Since the observed behavior of the MG is independent of
$s\geq 2$ we chose the minimal case $s=2$. Then introducing
$\sigma_i(t) = \mathrm{sign} [\delta p_i(t)]$, where $\delta
p_i(t) = p_i^1(t)-p_i^2(t)$ is the difference of points between
agent $i$ two strategies, equation (\ref{eq_points}) reads
\begin{equation}\label{eq_points1}
\delta p_i(t+1) = \delta p_i(t) - \vec{\xi}_i \cdot \vec{I}(t)\:
g[A(t)]
\end{equation}
and
\begin{equation}\label{eq_action1}
A(t) = \sum_{j=1}^N a_i(t) = \sum_{j=1}^N [\vec{\omega}_i + \sigma_i(t) \vec{\xi}_i ] \cdot \vec{I}(t)
\end{equation}
where $\vec{\omega}_i = (\vec{r}_i^{\:1} + \vec{r}_i^{\:2})/2$ and
$\vec{\xi}_i = (\vec{r}_i^{\:1} - \vec{r}_i^{\:2})/2$. With this
notation, $\beta_i(t) = (\sigma_i(t) + 3)/2$. Another unnecessary
complication is the binary payoff function $g(x) =
\mathrm{sign}(x)$ for the strategies. Thus, we focus on the linear
case $g(x) = x/D$ which allows for a simple treatment. In this
case we have
\begin{equation}\label{eq_action2}
\delta p_i(t+1) = \delta p_i(t) -\frac{1}{D}\ \vec{\xi}_i \cdot \vec{I}(t)\ \bigg\{\sum_{j=1}^N [\vec{\omega}_i + \sigma_i(t) \vec{\xi}_i ]\cdot \vec{I}(t) \bigg\}
\end{equation}

 and one is allowed to
neglect the fluctuations (law of large numbers)

In this form, the authors in \cite{cmzprl2000,mczphysa2000} argued
that if the information is just a random number then in the long
run each of the possible values of $\vec{I}(t)$ is visited with
equal probability. Thus, in the limit $N\to \infty$ and $D\to
\infty$ and making a temporal coarse-graining over time steps
$\tau = t/D$ that contain infinite number of updatings like
(\ref{eq_points1}), the right hand side of (\ref{eq_points1}) can
be replace by its mean value over time \cite{challetcontinous}:
\begin{equation}\label{eq_points3}
\delta p_i(t+\tau) = \delta p_i(t) - \ \vec{\xi}_i \cdot
\bigg\{\sum_{j=1}^N [\vec{\omega}_i + m_i(t) \:\vec{\xi}_i
]\bigg\},
\end{equation}
where $m_i(t) = \sum_{t'=t-\tau}^t\sigma_i(t')$ and we have
neglected fluctuations in favor of mean values according to the
law of large numbers. Finally, making the approximation that the
points at time $t$ are related to $m_i(t)$ through the soft
condition $m_i(t) = \tanh[\Gamma \delta p_i(t)]$ (where $\Gamma$
is a constant) we obtain the following dynamical equation in the
continuum approximation $\tau \to 0$
\begin{equation}\label{lyapunov1}
\frac{d m_i}{d\tau} = -2 \Gamma (1-m_i^2)\bigg[\sum_{j=1}^N \vec{\omega}_j\cdot \xi_i + \sum_j^N\vec{\xi}_i\cdot \vec{\xi}_j m_j\bigg].
\end{equation}
This can be easily written as a gradient descent dynamics
$dm_i/d\tau = -\Gamma(1-m_i^2) (\partial \Hcal / \partial m_i)$
where
\begin{equation}\label{lyapunov2}
\Hcal[\{m_i\}] = \frac{1}{2} \bigg[\sum_{i=1} (\vec{\omega}_i + \vec{\xi}_i m_i)\bigg]^2.
\end{equation}
Note that $\Hcal$ is a positive function of $m_i$ and then $\Hcal$
is a Lyapunov function of the dynamics given by (\ref{lyapunov1}).
Making the same type of temporal coarse-graining we get
\begin{equation}\label{sigma111}
\sigma^2 = \overline{\Hcal} + \sum_{i=1}^N
\overline{\vec{\xi}_i^{\:2}} (1-m_i^2).
\end{equation}
This equations implies that the dynamics of the agents tends to
minimize $\Hcal$ instead of $\sigma^2$ and thus the properties of
the MG are described by the ground states of $\Hcal$. It is easy
to see that $\overline{\Hcal}$ is related to the order parameter
$H$ at least when $H$ is very small, i.e. close to the phase
transition. Then, adaptative agents tend to minimize the
arbitrage/information rather than their collective losses
$\sigma^2$.

\begin{figure}
\begin{center}
\includegraphics[width=0.4\textheight]{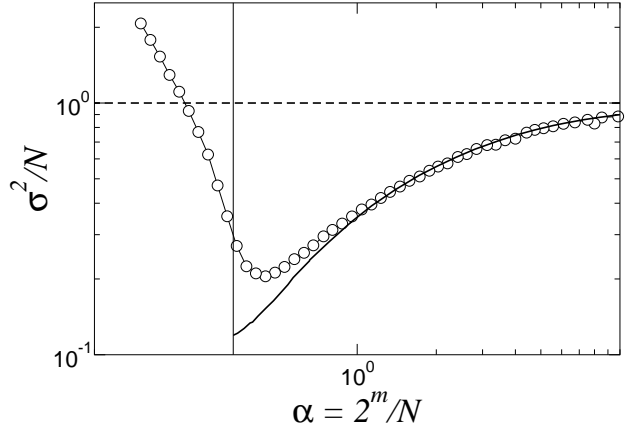}
\caption{Comparison of the results for the volatility in the
original MG (circles) with the solution obtained for equation
(\ref{sigma111}) through the evaluation of the minima of $\Hcal$}
\label{figtransition}
\end{center}
\end{figure}

Taking into account the random character of $\vec{\omega_i}$ and
$\vec{\xi}_i$, the Lyapunov function $\Hcal$ is reminiscent of of
the Sherrington-Kirkpatrick model \cite{sk} augmented by a random
field.  Ground state properties of $\overline{\Hcal}$ can be
analyzed using standard tools from statistical mechanics of
disordered systems. Averages over different realizations of the
disorder $\vec{\xi}_i$ and $\vec{\omega}_i$ are obtained through
the replica trick \cite{cmzprl2000,mczphysa2000}. Despite the
approximations made to get to equations (\ref{lyapunov1}) and
(\ref{lyapunov2}), the calculations showed that there is a phase
transition at $\alpha_c \simeq 0.33740$ and gave accurate
predictions for the values of $\sigma^2$ for $\alpha
> \alpha_c$ (see figure \ref{figtransition}): in fact
$\overline{\Hcal} = 0$ for $\alpha \leq \alpha_c$ while
$\overline{\Hcal} \neq 0$ when $\alpha \geq \alpha_c$. For $\alpha
\leq \alpha_c$ however, the solution given by exploring the ground
state of $\Hcal$ has some mathematical problems and fails to
reproduce the observed behavior.

The fact that the game for $\alpha > \alpha_c$ is described by the
minimization of $\Hcal$ shed some light also about the real nature
of the MG: first of all, the dynamics tend to minimize a Lyapunov
function which depends on the history of agents decisions instead
of agents instantaneous decisions. And second, the function which
agents tend to minimize has some disorder which is encoded in the
``quenched'' patterns $\vec{\xi}_i$ and $\vec{\omega}_i$. This
last feature of the MG resembles the dynamics of attractor neural
networks (ANN) \cite{amit} where $D$ patterns are stored through
the Hebbian learning rule and the recurrent network retrieve the
information stored in the neurons. The control parameter $\alpha =
D/N$ measures then the ratio between the number of stored patterns
$D$ and the number of neurons $N$. In general, information
retrieval in ANN is possible only under some conditions about the
dynamics and $\alpha$. This comparison is very appealing, since in
principle agents in the MG would like to retrieve time patterns or
other agents strategies to achieve a better personal record. Thus,
is MG just an ANN of neurons trying to retrieve other agents'
strategies? We will complete this analogy in the following
sections.

\section{No news, good news}\label{sec_news}
The irrelevance of the information contained in the sequence of
the winning groups found in section \ref{sec_information} poses a
new question: how can it be that changing the model from being a
closed system with endogenous information to an open one with
exogenous information does not change its aggregate behavior? As
we saw, the reason for that is that the aggregate behavior of the
MG is encoded in the response of the system to the input
information. Averaging this response over all possible values of
the information will give us the mean response of the system and
then we can update all strategies'  points concurrently. For
example, we found in section \ref{sec_maximizing} that
coarse-graining the dynamics of the MG over many time steps leads
to a coupled set of equations for $p_i^{\alpha}(t)$ in which the
information $\vec{I}(t)$ is averaged out. The information is then
simply a go-between quantity that connects agents strategies
dynamics and in principle it is disposable.

In order to get the averaged dynamics of the strategies for all
possible values of $\alpha$ and not just for $\alpha > \alpha_c$
as in the previous section, several techniques were used
\cite{moro,garrahan1,heimel,challetcontinous,sherrington3}. Here
we outline one of them, which is by means of the Kramers-Moyal
expansion \cite{gardiner}. This is a general technique to obtain
diffusion approximations of stochastic processes. Specifically,
equations (\ref{eq_action2}) define an stochastic process than can
be approximated by a drift together with a diffusion in the space
of strategies' points. In this case, due to the fact that the
random process $\vec{I}(t)$ appears twice in equation
(\ref{eq_action2}), only the drift term remains non-zero in the
Kramers-Moyal approximation which yields
\begin{equation}\label{eq_action3}
\delta p_i(t+1) = \delta p_i(t) - \vec{\xi}_j \cdot
\bigg\{\sum_{j=1}^N\vec{\omega}_j + \sigma_j(t)
\vec{\xi}_j\bigg\}.
\end{equation}
Comparing equation (\ref{eq_action3}) with (\ref{eq_action2}) we
observe that information has been averaged out in favor of an
effective coupling between agents' strategies, as we expected.
Thus, rather than allowing the strategy payoff valuations to be
changed at each round, only the accumulated effect on a large
number of market decisions is used to change an agent's strategy
payoff valuations. We regard equation (\ref{eq_action3}) as the
equivalent of what in the neural network literature would be
called the {\em batch} version of the conventional {\em on-line}
MG \cite{heimel}.

\begin{figure}
\begin{center}
\includegraphics[width=0.45\textwidth]{fig-moro}\ \ \
\includegraphics[width=0.45\textwidth]{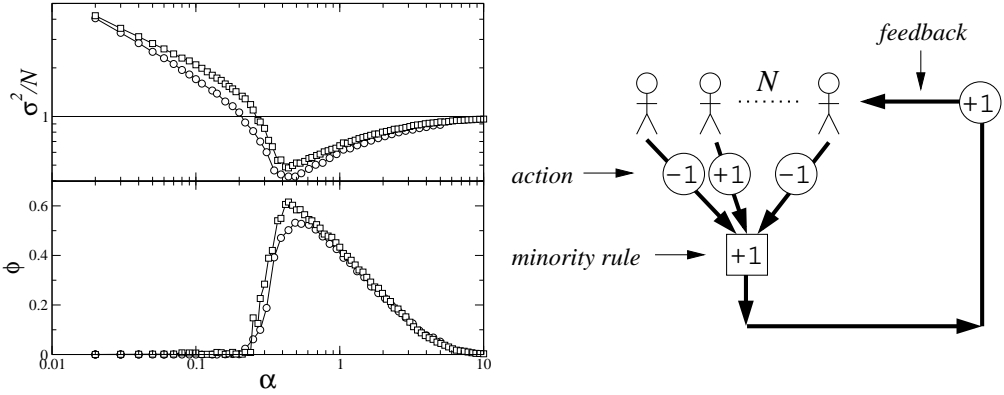}
\caption{{\em Left:} Results of the MG using equations
(\ref{eq_action3}) (squares), compared with simulations of the
original MG (circles). Parameters used are $N=100$ and $s=2$. {\em
Right:} Cartoon of the MG model after averaging out the
information: feedback performed by the agents to keep track of
their strategies performance provides an effective interaction in
the system. As in Fig.\ \ref{fig_tabla}, solid thick lines mimic
the flow of information in the system.} \label{figcartoon2}
\end{center}
\end{figure}

Equations (\ref{eq_action3}) are the simplest version of the
dynamics to reproduce the observed behavior in the MG for all
values of $\alpha$, as shown in figure \ref{figcartoon2}. Since
fluctuations due to the information are averaged out, there is a
small quantitative difference in the aggregates (like $\sigma^2$),
but its main features are qualitatively conserved. Within this
approximation, the volatility is given by
\begin{equation}\label{volatility_batch}
\sigma^2 = \overline{\Omega} + 2\sum_{i=1}^N \overline{h_i \mean{\sigma_i(t)}}-\sum_{i,j=1}^N \overline{J_{ij} \mean{\sigma_i(t)\sigma_j(t)}}
\end{equation}
where $\Omega = \sum_{j=1}^N \vec{\omega_i}\cdot\vec{\omega_j} / D
$, $h_i = \sum_{j=1}^N \vec{\omega}_j \cdot \vec{\xi}_i /D $ and
\begin{equation}\label{hebbian}
J_{ij} =  -\frac{1}{D}\vec{\xi}_i \cdot \vec{\xi}_j.
\end{equation}

Although there are some quantitative differences between the
original MG and the effective dynamics in the strategy space given
by equations (\ref{eq_action2}), the approximations given by
(\ref{eq_action3}) reconciliates the model with its endogenous
nature. On the other hand, note that equations (\ref{eq_action3})
look like equations (\ref{eq_points3}) with the accumulated agent
temporal pattern $m_i(t)$ instead of the instantaneous
$\sigma_i(t)$. In fact, equations (\ref{eq_action3}) can be
written in the continuum limit like
\begin{equation}\label{hamiltonian2}
\frac{d (\delta p_i)}{dt} = -\frac{\partial \Hcal[\{\sigma_i\}]}{\partial \sigma_i}
\end{equation}
where $\Hcal[\{\sigma_i\}]$ is given by equation (\ref{lyapunov2}) with $\sigma_i(t)$ instead of $m_i(t)$. This time, $\Hcal$ with $\sigma$'s instead of $m$'s is not a Lyapunov function of (\ref{eq_action3}) in general, since the gradient is with respect to $\sigma$, not $p$. But, as we saw in section \ref{sec_maximizing}, $\Hcal$ is a Lyapunov function for $\alpha > \alpha_c$ for $m_i = \mean{\sigma_i(t)}$ \cite{moro,sherrington3}.

Since equations (\ref{eq_action3}) are the simplest dynamics of
the MG, we are in the position to single out the main features
responsible for the behavior of the MG. To this end, let us
rewrite equations (\ref{eq_action3}) the following way:
\begin{equation}\label{mgann}
\sigma_i(t+1) = \mathrm{sign}\left[\frac{\delta p_i(0)}{t} - \Omega_i + \sum_{j=1}^N J_{ij} x_i(t)\right]
\end{equation}
where $x_i(t) = \frac{1}{t}\sum_{\tau=0}^t \sigma(\tau)$ is the
time average of agent $i$ actions over time. If we assume that
$\delta p_i(0) =0$, equation (\ref{mgann}) looks like an attractor
neural network \cite{amit} (specifically a Hopfield model), where
the patterns $\vec{\xi}$ have been learned through the {\em
anti-Hebbian} rule (\ref{hebbian}) instead of the usual Hebbian
rule $J_{ij} = \vec{\xi}_i \cdot \vec{\xi}_j$, and each neuron is
subject to external biases $\Omega_i$. In ANNs anti-Hebbian rule
is studied in models of paramagnetic {\em unlearning}
\cite{hopfield,nokura}, the process by which spurious states from
the space of possible configurations of $\sigma_i(t)$ are removed
since they do not correspond to any stored pattern\footnote{Some
biologists have suggested that this unlearning procedure
correspond to REM sleep, which is widely observed among mammals
\cite{hopfield}.}; the opposite sign in the anti-Hebbian rule
hinders rather that assists information retrieval in ANN. In the
MG, the origin of the anti-Hebbian rule is different: it comes
through the minority rule, which forces the behavior of agents in
the long run to be different to any other agents behavior.

More generally we can look at systems like
\begin{equation}\label{mgann1}
\sigma_i(t+1) = \mathrm{sign}\left[- \Omega_i + \sum_{j=1}^N J_{ij} x_i(t)\right]
\end{equation}
where $x_i(t) = \frac{1}{M} \sum_{\tau = 0}^{M} \sigma_i(t-\tau)$
is the time average of agent $i$ behavior over the last $M$ time
steps. This kind of time delayed ANN were considered as possible
candidates of models for storing and generating time sequences of
patterns \cite{kinzel,marcus}. In the MG, the time delayed
dynamics is not set up to retrieve or create any particular time
pattern but rather comes from the inductive reasoning process,
which keeps record of the performance of the strategies along the
game.

\begin{figure}
\begin{center}
\includegraphics[width=0.666\textwidth,clip=]{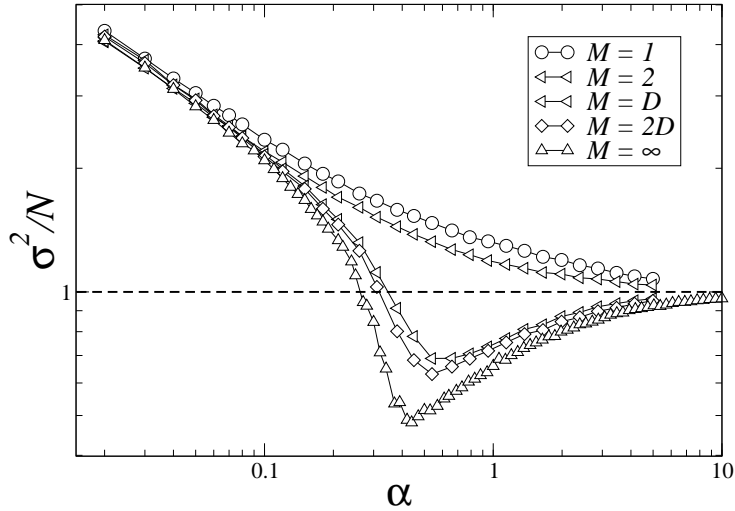}
\caption{Volatility obtained through simulations of equations
(\ref{mgann1}) for different values of $M$ with $N=100$.}
\label{fig-mm}
\end{center}
\end{figure}

Thus, equations (\ref{mgann1}) define a new type of ANN whose main
ingredients are:
\begin{itemize}
\item $D$ heterogenous patterns are stored through the
anti-Hebbian rule (\ref{hebbian}) due to the minority rule, \item
each neuron reacts to an external field given by $\Omega_i$, and
\item dynamics is time delayed over the last $M$ steps because of
the adaptation process,
\end{itemize}
which are necessary to reproduced the observed behavior of the MG.
Finally one may ask whether is it relevant to keep the whole
historic performance of each strategy, i.e., $M \to \infty$ or
just the last $M < \infty$ time steps\footnote{A similar model
(named time horizon MG), was considered in \cite{johnsonthorizon}
in which, in the original MG model, $p_i^{\alpha}(t)$ are just the
points of strategy $\alpha$ of agent $i$ for the last $M$ time
steps.}. In figure \ref{fig-mm} we see that if only the last $M$
strategy performances are kept with $M$ small, then the system
never attains better result than the random case. Actually the
cases $M=1,2$ reproduces quite accurately the high volatility
region $\alpha \ll \alpha_c$ which means that in this phase,
agents react only to the last winning group. For $\alpha \geq
\alpha_c$ keeping the record of strategies over large time periods
$M \gg D$ is the only way for the agents to achieve a
better-than-random solution to the problem.

Finally, it is appealing to rewrite equations (\ref{mgann}) like a
learning process \cite{amit}. After all, inductive reasoning in
the MG is the way agents learn how to act in the game. If the time
dependence in (\ref{mgann1}) is considered as a learning process
or training of a set of $N$ interacting perceptrons, then the
learning process is given by
\begin{eqnarray}
\sigma_{i}(t+1)&=&\mathrm{sign}[-\Omega_i + \sum\tilde J_{ij}(t) \sigma_j] \label{task} \\
\tilde J_{ij} (t)&=&\frac{t}{t+1}\tilde J_{ij} \sigma_j(t+1) \sigma_j(t)+ \frac{1}{t+1}J_{ij}
\label{learning}
\end{eqnarray}
This type of interacting $N$ perceptrons (one for each agent) have
been considered recently in the context of the MG \cite{kinzel0},
although with a different learning rule for the perceptrons and
with homogeneous agents (i.e. $\vec{\xi}_i = \vec{\xi}_j,
\vec{\omega}_i = \vec{\omega}_j$). However, as we can see in
(\ref{learning}), the MG has its unique learning process which
comes through adaptation of agents along iteration of the game.

In summary, after clearing up the MG model we ended up with a set
of heterogenous agents (or perceptrons) which are learning through
the adaptation process (\ref{learning}) the minority task given by
(\ref{task}). It would be interesting to extend the training of
$N$ perceptrons given by (\ref{learning}) and to consider the
general problem of learning in a set of interacting $N$
perceptrons under the basic MG features: heterogeneity, minority
rule and time delayed interactions.

\section{Dynamics matter}\label{sec_dynamics}

The success of the solution obtained using the replica trick
\cite{cmzprl2000,mczphysa2000} and outlined in section
\ref{sec_maximizing}, established the concept that inductive
reasoning of $N$ interacting agents in the MG is described by the
minimization of a global function, the response $\Hcal$. Thus, the
MG seems to be similar to an equilibrium problem of disordered
systems. But a simple numerical experiment shows that this is not
the case \cite{moro}: Suppose that agents are confident about one
of their strategies and they start by giving it a period of grace,
i.e. an initial number of points different from zero. This means
that $p_i^\alpha(0) \neq 0$ for some $\alpha$  in equation
(\ref{eq_points}). Simulations in figure \ref{inicond} show that
this tiny modification of the MG model has a profound impact in
the results for $\alpha \leq \alpha_c$. Specifically, the high
volatility region is lost and a region in which volatility is very
small emerges. Moreover, the steady state volatility for $\alpha
\leq \alpha_c$ depends on the initial condition $p_i^\alpha(0)$
\cite{coolen_rev,moro,heimel}\footnote{Due to the dependence on
the initial condition for $\alpha < \alpha_c$, experiments in
which the number of agents $N$ or the memory $D$ are changed
quasi-statically lead to hysteresis behavior in the MG
\cite{sherrington3}, again demonstrating the non-equilibrium
behavior in the region $\alpha < \alpha_c$}.

The fact that the steady state reached by agents through
adaptation depends on the initial condition for $\alpha \leq
\alpha_c$ means that this region is non-ergodic and that the
observed change of behavior at $\alpha = \alpha_c$ is due to a
non-equilibrium phase transition instead of an equilibrium
transition. A precise mathematical description of this was reached
through the generating functional methods of De Dominicis
\cite{dedominicis} by A.\ C.\ C.\ Coolen and coworkers
\cite{coolen1,coolen2,coolen_rev,heimel}. This technique allows to
carry out the average over the quenched disorder in the system
into the dynamical equations directly, instead of performing it in
the equilibrium partition function. The outcome of this averaging
is a set of closed equations for the correlation $C(t,t')$ and
response $R(t,t')$ functions of the variables involved in the
dynamical equations [$p_i^\alpha(t)$ and $\sigma_i(t)$] at
different times. This system of dynamical equations for $C(t,t')$
and $R(t,t')$ is difficult to solve in general, but some
analytical progress can be made if one looks at two particular
situations:

\begin{figure}
\begin{center}
\includegraphics[width=0.66\textwidth,clip=]{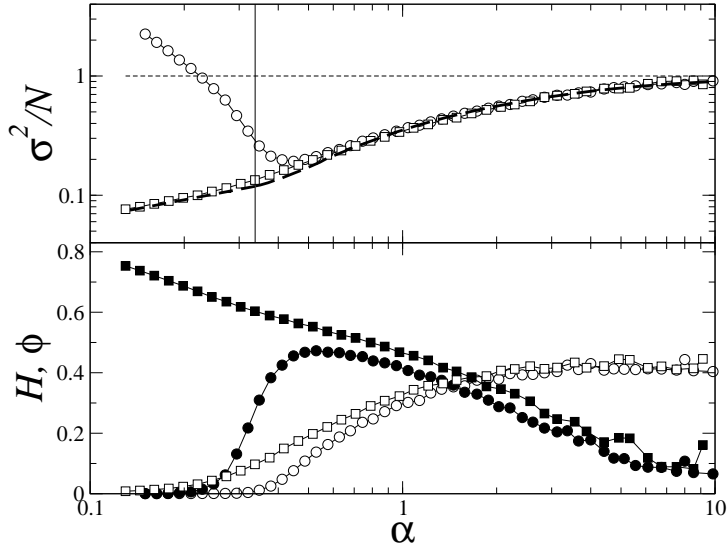}
\caption{Simulations of the original MG model (with endogenous
information) for different initial conditions and $m=7$, $s=2$ and
$p_i^1(0) = 0$ (circles) and $p_i^1(0) = 100$ (squares). {\em
Upper panel:} dashed lines correspond to the stationary solution
for $\alpha > \alpha_c$ and $\sigma^2 /N \sim \alpha^{1/2}$ for
$\alpha \leq \alpha_c$. {\em Lower panel:} full symbols correspond
to $H$ while open symbols refer to $\phi$. } \label{inicond}
\end{center}
\end{figure}

\begin{itemize}
\item For instance, one can investigate asymptotic stationary
solutions, i.e. when $C(t,t) = C(t-t')$ and $R(t,t') = R(t-t')$.
and assume that the system has not anomalous response, i.e. that
fluctuations around the stationary state are damped in finite time
and the susceptibility is finite. In this case, the results
obtained in section \ref{sec_maximizing} through the evaluation of
the minima of $\Hcal$ are recovered. Moreover, this stationary
solution predicts self-consistently when it breaks down: the
susceptibility of the solution diverges at $\alpha = \alpha_c$ and
the assumption of stationary in the solution is only valid for
$\alpha > \alpha_c$. \item For $\alpha \leq \alpha_c$ one can
investigate the first time steps of the correlation and response
function dynamics. It is found that the volatility depends
strongly on the initial conditions: for $s=2$ and depending on
$\delta p_i(0)$ a transition from high volatility $\sigma^2/N \sim
\alpha^{-1}$ to low volatility $\sigma^2/N \sim \alpha^{1/2}$ is
found (see Fig.\ \ref{inicond}).
\end{itemize}

These results point out that the phase transition found in the MG
is a non-equilibrium transition and that only for $\alpha >
\alpha_c$ the adaptation dynamics is relaxing and ergodic: for any
initial condition a final unique steady state is reached which is
described by the minima of $\Hcal$. For $\alpha \leq \alpha_c$ the
system is non-ergodic, since the system asymptotic behavior
depends strongly on the initial conditions. After all, the MG is
posed without requiring detailed balance or any kind of
Fluctuation-Dissipation theorem about its dynamics, so there is in
principle no reason to expect that the MG should have the usual
relaxing and ergodic dynamics of other models in statistical
mechanics. Finally, it is interesting to note \cite{coolen_rev}
than ergodicity in the MG is not broken at the macroscopic level
as in replica symmetry breaking solutions in spin-glass; it is
done at the microscopic level since the dynamical rules do not
fulfill any type of detailed balance.

The generating functional analysis proved to be the right tool to
study the MG since its focus is on the dynamics of disordered
systems, specially on spin glasses. Since the existence of a
partition function is not needed, complicated cases like steady
states with non-equilibrium fluctuation-dissipation relationships
or violations of detailed balance can be tackled within this
formalism. Indeed, the MG model features do not root in physical
considerations so there is no reason {\em a priori} to expect
detailed balance or fluctuation-dissipation relationships.

\section{Outlook}\label{sec_outlook}
\begin{flushright}
\begin{minipage}{8cm}
\scriptsize The land of statistical physics is broad, with many
dales, hills, valleys and peaks to explore
that are of relevance to the real world and to our ways of thinking about it.\\
{\em Michael E.\ Fisher in Rev. Mod. Phys. {\bf 70}, 653 (1998)}.
\end{minipage}
\end{flushright}
Through this journey we have learned that adaptation or inductive
reasoning as that of the MG model leads to a system of many
interacting and heterogeneous degrees of freedom whose dynamics is
not local in time. The system may or may not be ergodic and,
generally speaking, equilibrium statistical mechanics should be
abandoned. However there are some similarities with statistical
mechanics of disordered systems and some of the tools and concepts
of disordered system do apply to the MG. The MG is just a model of
adaptation of $N$ agents which can be generalized to other types
of learning as we saw in section \ref{sec_news}. For those types
of dynamical problems, the suitable tool from statistical physics
is the generating functional approach we mentioned in section
\ref{sec_dynamics} which gives not only information about the
properties of the system when a steady state is reached, but also
about the dynamical nature of the phase transition. In fact, the
generating function approach is also useful in different
modifications of the MG \cite{coolen1,coolen2,coolen_rev,galla}.

Adaptation turned out to be a good way to solve the MG problem for
$\alpha > \alpha_c$. But it might not be the best way to do it in
the region $\alpha \leq \alpha_c$. As a matter of fact, simple
modifications of the original MG model like taking a non-zero
initial condition or what is called the Thermal Minority Game
\cite{cavagna3,cavagna4,challetcomment}, in which agent's
decisions are taken with some degree of stochasticity (instead of
using their best one always), lead to better solutions ($\sigma^2/
N \ll 1$) of the MG problem than the original model.

However, part of the success of the application of analytical
techniques in the MG is due to its fixed rules. For example, in
real life we do not expect to come across a game in which players
have the same space of strategies, or in which the initial
condition for all of them is that in which none of their
strategies is preferred. Relaxing some of these rules in the MG
does not modify its behavior; unfortunately, others do. During the
last years we have witnessed a burst of spin-off models whose core
is given by the MG \cite{challetwww}. Of particular importance are
those in which refinements of the model were towards the
description of real markets
\cite{challetmarket,bouchaud1,bouchaud2,marsili_review} or real
problems in ecology \cite{macara}. The validity of the MG as a
behavioral model of agents trying to spot patterns in aggregate
outcomes has been even tested through experiments with people
\cite{experiment1,experiment2} showing that indeed people tend to
coordinate despite its selfish nature in the game.

On the other hand, there has been some extensions of the MG which
has been used in the context of time series analysis and
prediction by doing {\em reverse engineering}: since the MG is, as
we saw in section \ref{sec_news}, a precise learning rule, why not
training a set of $N$ agents with a real life time series? In that
case, agents strategies are not fixed at the beginning of the game
but rather, they are chosen to reproduced the given time series.
This has been applied to financial times series in order to
forecast future  price movements \cite{johnsonprediction}.

The virtue of the MG relies on the incorporation of the minority
rule and agents' heterogeneity in the model as well as an
agent-based approach to the solution of a given problem. In this
regard, during the last years there has been a lot of attention in
the solution of general problems using multi-agent models
\cite{mass,mass2}. In fact coupled layers of perceptrons are just
a simple example of this approach \cite{amit}. The general
principles of those multi-agent systems is that agents
individually have incomplete information or capabilities for
solving the problem but can reach a solution by interacting among
them. The success of understanding of the MG using tools from
statistical mechanics encourages to tackle more general
heterogeneous multi-agent systems problems within the theory of
disordered systems in condensed matter.

\bigskip

{\bf Acknowledgments} I am grateful to my collaborators J.\ P.\
Garrahan and D.\ Sherrington at University of Oxford where most
part of our work was done. I benefited also from discussions with
J.-P.\ Bouchaud, M.\ A.\ R.\ de Cara, A.\ Cavagna, D.\ Challet,
A.\ C.\ C.\ Coolen, T.\ Galla, F.\ Guinea, N.\ F.\ Johnson, G.\
Lythe, M.\ Marsili and A.\ S\'anchez. This work was partially
supported by EPSRC (United Kingdom), European Commission and
Ministerio de Ciencia y Tecnolog\'{\i}a (Spain).


\begin{thebibliography}{99}
\bibitem{amit} D.\ J.\ Amit, {\em Modeling brain function: the world of Attractor Neural Networks}, (Cambridge University Press, Cambridge, 1989).
\bibitem{anderson1} {\em The Economy as an Evolving Complex System}, edited by P.\ W.\ Anderson, K.\ Arrow and D.\ Pines (Addison-Wesley, Redwood City, CA, 1988).
\bibitem{arthur94} W.\ B.\ Arthur., Am.\ Econ.\ Assoc.\ Papers and Proc.\ {\bf 84}, 406 (1994).
\bibitem{experiment1} G.\ Bottazi and G.\ Devetag, Physica A {\bf
324}, 124 (2003).
\bibitem{cavagna1} A.\ Cavagna, Phys. Rev. E {\bf 59} R3783 (1999).
\bibitem{cavagna2} A.\ Cavagna, Phys.\ Rev.\ Lett.\ {\bf 84}, 1058 (2000).
\bibitem{cavagna3} A.\ Cavagna, J.\ P.\ Garrahan, I.\ Giardina, and D.\ Sherrington, Phys.\ Rev.\ Lett.\ {\bf 83}, 4429 (1999).
\bibitem{cavagna4} A.\ Cavagna, J.\ P.\ Garrahan, I.\ Giardina, and D.\ Sherrington, Phys.\ Rev.\ Lett.\ {\bf 85}, 5009 (1999).
\bibitem{macara1} M.\ A.\ R.\ de Cara, O.\ Pla, and F.\ Guinea, Eur.\ Phys.\ J.\ B {\bf 10}, 187 (1999)
\bibitem{macara} M.\ A.\ R.\ de Cara, PhD Thesis (2003).
\bibitem{challetwww} D.\ Challet maintains a personal commented collection of papers and preprints on the MG at the web page:  http://www.unifr.ch/econophysics/minority/.
\bibitem{challet1997} D.\ Challet and Y.-C. Zhang, Physica A {\bf 246}, 407
  (1997).
\bibitem{challet1998} D.\ Challet and Y.-C.\ Zhang, Physica A {\bf 256} 514 (1998).
\bibitem{challetmarsilipre99} D.\ Challet and M.\ Marsili, Phys.\ Rev.\ E (1999).
\bibitem{cmzprl2000} D.\ Challet, M.\ Marsili and R.\ Zecchina, Phys.\ Rev.\ Lett.\ {\bf 84}, 1824 (2000).
\bibitem{cmzphysa2000} D.\ Challet, M.\ Marsili and Y.-C.\ Zhang, Physica A {\bf 276}, 284 (2000).
\bibitem{challetmemory} D.\ Challet and M.\ Marsili, Phys.\ Rev.\ E {\bf 62}, 1862 (2000).
\bibitem{challetcomment} D.\ Challet, M.\ Marsili and R.\ Zecchina, Phys.\ Rev.\ Lett.\ {\bf 85}, 5008 (2000).
\bibitem{challetmarket} D.\ Challet, M.\ Marsili and Y.-C.\ Zhang, Physica A {\bf 294}, 514 (2001).
\bibitem{coolen1} A.\ C.\ C.\ Coolen, J.\ A.\ Heimel and D.\ Sherrington, Phys.\ Rev.\ E {\bf 65} 016126 (2001)
\bibitem{coolen2} A.\ C.\ C.\ Coolen and  J.\ A.\ Heimel, J.\ Phys. A: Math. Gen.\ {\bf 34} 10783 (2001).
\bibitem{coolen_rev} A.\ C.\ C.\ Coolen, {\em Non-equilibrium statistical mechanics of Minority Games}, in Proceedings of Cergy 2002 Conference (2002).
\bibitem{deangelis} {\em Individual-based models and approaches in ecology: populations, communities, and ecosystems}, edited by D. DeAngelis and L.\ Gross, (Chapman \& Hall, New York, 1992).
\bibitem{dedominicis} C.\ De Dominicis, Phys.\ Rev.\ B {\bf 18}, 4913 (1978).
\bibitem{gametheory} D.\ Fudenberg and J.\ Tirole, {\em Game Theory} (MIT Press, Cambridge, 1991).
\bibitem{galla} T.\ Galla, A.\ C.\ C.\ Coolen, and D.\
Sherrington, J.\ Phys.\ A. (in press) (2003)
\bibitem{gardiner} C. W. Gardiner, {\em Handbook of Stochastic Methods},
(Springer, Berlin, 1996).
\bibitem{moro} J.\ P.\ Garrahan, E.\ Moro and D.\ Sherrington, Phys.\ Rev.\ E {\bf 62}, R9 (2000).
\bibitem{garrahan1} J.\ P.\ Garrahan, E.\ Moro and D.\ Sherrington, Quant. Finance {\bf 1}, 246
(2001).
\bibitem{bouchaud1} I.\ Giardina, J.-P.\ Bouchaud and M.\
M\'ezard, Physica A {\bf 299} 28 (2001).
\bibitem{bouchaud2} I.\ Giardina and J.-P.\ Bouchaud, Physica A
{\bf 324}, 6 (2003).
\bibitem{johnsoncrowd} M.\ Hart, P.\ Jefferies, N.\ F.\ Johnson, and P.\ M.\ Hui, Physica A {\bf 298}, 537 (2001).
\bibitem{johnsonthorizon} M.\ Hart, P.\ Jefferies and N.\ F.\ Johnson, Physica A {\bf 311}, 275 (2002).
\bibitem{heimel} J.\ A.\ Heimel and A.\ C.\ C.\ Coolen, Phys.\ Rev.\ E {\bf 63} 056121 (2001).
\bibitem{hopfield} J.\ J.\ Hopfield, D.\ I.\ Feinstein, and R.\ G.\ Palmer, Nature {\bf 304}, 158 (1983).
\bibitem{huberman} B.\ Huberman, R.\ Lukose, Science {\bf 277}, 535 (1997).
\bibitem{dhulstrodgers99} R.\ D'hulst and G.\ J.\ Rodgers, Physica A {\bf 270} 222 (1999).
\bibitem{johnson1} N.\ F.\ Johnson et al. PHysica A {\bf 256} 230 (1998).
\bibitem{johnsonprediction} N.\ F.\ Johnson, D.\ Lamper, P.\ Jefferies, M.\ L.\ Hart, S.\ Howison, Physica A {\bf 299}, 222 (2001).
\bibitem{kinzel0} W.\ Kinzel, R.\ Metzler and I.\ Kanter, J.\ Phys.\ A: Math.\ Gen.\ {\bf 33}, L141 (2000).
\bibitem{kinzel} W.\ Kinzel, R.\ Metzler and I.\ Kanter, Physica A {\bf 302}, 44 (2001).
\bibitem{kosfeld} M.\ Kosfeld, {Individual decision-making and Social Interaction}, CentER Dissertation Series, Vol. 55 (1999).
\bibitem{lee} C.-Y.\ Lee, Phys. Rev. E {\bf 64}, 015102 (2001).
\bibitem{li2000} Y.\ Li, A. VanDeemen and R.\ Savit, Physica A {\bf 284}, 461 (2000)
\bibitem{manucaphysa2000} R.\ Manuca, Y.\ Li, R.\ Riolo and R.\ Savit, Physica A, {\bf 282}, 559 (2000).
\bibitem{marcus} C.\ M.\ Marcus and R.\ M.\ Westervelt, Phys.\ Rev.\ A {\bf 42}, 2410 (1996).
\bibitem{mczphysa2000} M.\ Marsili, D.\ Challet and R.\ Zecchina, Physica A {\bf 280}, 522 (2000).
\bibitem{marsili_review} M.\ Marsili, {\em Toy models of markets
with heterogeneous interacting agents} in {\em Economics With
Heterogeneous Interacting Agents} A.\ Kirman and J.-B.\ Zimmerman
(editors), Lecture Notes in Economics and Mathematical Systems
Vol. 503, page 161 (Springer-Verlag, 2001).
\bibitem{challetcontinous} M.\ Marsili and D.\ Challet, Phys.\ Rev.\ E {\bf 64}, 056138 (2001).
\bibitem{nagel1} K. Nagel, S. Rasmussen, C.\ Barrett in {\em Self-organization of complex structures: from individual to collective dynamics} edited by F.\ Schweitzer (Gordon and Breach, London, 1997), p.\ 579.
\bibitem{nokura} K.\ Nokura, J.\ Phys.\ A: Math.\ Gen.\ {\bf 31} 7447 (1998).
\bibitem{experiment2} T.\ P\l atkowski and M.\ Ramsza, Physica A
{\bf 323} 726 (2003).
\bibitem{savitprl1999} R.\ Savit, R.\ Manuca and R.\ Riolo, Phys. Rev. Lett. {\bf 82}, 2203 (1999).
\bibitem{savitprl2000} R.\ Savit, Phys.\ Rev.\ Lett.\ {\bf 84}, 1059 (2000).
\bibitem{schelling} T.\ C.\ Schelling, {\em Micromotives and Macrobehavior} (W.\ W.\ Norton, New York, 1978).
\bibitem{sk} D.\ Sherrington and S.\ Kirkpatrick, Phys.\ Rev.\
Lett.\ {\bf 35}, 1972 (1975).
\bibitem{sherrington99} D. Sherrington {\em Spin glasses}
in {\em Physics of novel materials} edited by M.P. Das (World
Scientific, Singapore, 1999), p.\ 146.
\bibitem{sherrington3} D.\ Sherrington, E.\ Moro and J.\ P.\ Garrahan, Physica A {\bf 311}, 527 (2002).
\bibitem{mass} G. Weiss, {\em Multiagent Systems: A Modern Approach to Distributed Artificial Intelligence} (MIT Press, 2000).
\bibitem{mass2} M.\ Wooldridge, {\em An Introduction to Multiagent Systems}, (John Wiley \& Sons Chichester, 2002).
\bibitem{young} A.P. Young (editor) {\em Spin glasses and random fields} (World Scientific,
Singapore, 1998).
\bibitem{hpyoung} H.\ P.\  Young, {\em Individual strategy and social structure} (Princeton University Press, Princeton, 1998).
\bibitem{zhang} Y.-C.\ Zhang, Europhys. News {\bf 29}, 51 (1998).
\bibitem{zheng} D.\ Zheng and B.-H.\ Wang, Physica A {\bf 301}, 560 (2001)




\end{thebibliography}
\end{document}